\documentclass[conference]{IEEEtran}

\usepackage{amssymb}
\usepackage{amsmath}

\usepackage{amsthm,theorem}
\usepackage{amssymb}
\usepackage{graphicx,tikz}
\usetikzlibrary{arrows,automata}

\usepackage{array,cite}
\usepackage{subfigure,epsfig}

\title{Relaying in Diffusion-Based Molecular Communication}
\author{  Arash Einolghozati, Mohsen Sardari, Faramarz Fekri\\
School of Electrical and Computer Engineering\\
Georgia Institute of Technology, Atlanta, GA 30332\\
\texttt{Email:}\{einolghozati, mohsen.sardari, fekri\}@ece.gatech.edu
\thanks{This material is based upon work supported by the National Science Foundation under Grant No. CNS-111094}
}

\begin{document}
\IEEEoverridecommandlockouts
\maketitle

\begin{abstract}
THIS PAPER IS ELIGIBLE FOR THE STUDENT PAPER AWARD. 

Molecular communication between biological entities is a new paradigm in communications.
Recently, we studied molecular communication between two nodes formed from synthetic bacteria. Due to high randomness in behavior of bacteria, we used a population of them in each node. The reliability of such communication systems depends on both the maximum  concentration of molecules that a transmitter node is able to produce at the receiver node as well as the number of bacteria in each nodes. This maximum concentration of molecules falls with distance which makes the communication to the far nodes nearly impossible. In order to alleviate this problem, in this paper, we propose to use a molecular relaying node. The relay node can resend the message either by the different or the same type of molecules as the original signal from the transmitter. We study two scenarios of relaying. In the first scenario, the relay node simply senses the received concentration and forwards it to the receiver. We show that this sense and forward scenario, depending on the type of molecules used for relaying, results in either increasing the range of concentration of molecules at the receiver or increasing the effective number of bacteria in the receiver node. For both cases of sense and forward relaying, we obtain the resulting improvement in channel capacity. We conclude that multi-type molecular relaying outperforms the single-type relaying. In the second scenario, we study the decode and forward relaying for the M-ary signaling scheme. We show that this relaying strategy increases the reliability of M-ary communication significantly.
\end{abstract}


\section{Introduction}
\label{sec:intro}
 
Recent advances in synthetic biology have encouraged using engineered bacteria as the basic components of a network. Molecular communication potentially offers an alternative to wired communications while the applications have yet to be developed. Bio-compatible environments like human-body are among the most promising scenarios for molecular communication. The widely studied example of molecular communication is the one observed among bacteria. It has been understood that bacteria use concentration of small molecules to signal each other in a process called Quorum Sensing~\cite{Bassler1999}. This process enables bacteria to measure the density of their population and hence, behave collaboratively for  performing the tasks that would be impossible otherwise.  As such, molecular communication has recently received a great deal of attention from several researchers both in biology and communication. The challenges of the molecular communication as a new frontier in the wireless communication are discussed in~\cite{MianRose,Monaco,Wireless_Molecular, ISIT2011_Arash, ITW2011_Arash}.

Throughout our work, we focus on diffusion-based molecular communication where the information is embedded in the concentration of molecules. This is in contrast to the other frontier of molecular communication which the information is encoded in the timing of molecules~\cite{Rose2012,eckford_IT}. In those scenarios, the individual behavior of molecules is the determinant of the decoded signal. In~\cite{INFOCOM2012_Arash}, we introduced the structure of nodes in a molecular communication network. Each node is  comprised of a population of bacteria inside a chamber. The rationale is that the individual behavior of bacteria contains huge randomness and hence, we use a population of them in order to build reliable nodes out of unreliable bacteria. Therefore, the collective behavior of bacteria influences the node output as a transmitter as well as a receiver. Such networks can be used for sensing the environment for the density of a particular particle and sending that information reliably to the destination. In~\cite{ISIT2012_Arash}, we studied the information sensing capacity of such a node and learned that the sensing capacity increases by using a larger number of bacteria in a node. In~\cite{ITW2012_Arash}, we showed how the reliable communication between two nodes is achieved and obtained the capacity and reliability for such a system. We also showed that the achievable rates of information increases with increasing the maximum range of the concentration of molecules induced at the receiver and a larger number of bacteria in the nodes.

In molecular communication, attenuation of the molecular concentration as it travels in the environment via the diffusion process is a major problem. Using the steady-state of the concentration of molecules at the receiver tries to mitigate this problem. This is due to the fact that at the steady state, the concentration of molecules is inversely proportional to the distance between the transmitter and the receiver. However, communication to the long distances still remains a challenge especially whenever other types of molecules are present at the environment, making the sensing of the signal molecules more difficult in the lower concentration regime. On the other hand, we are constrained to use as few number of bacteria as possible inside the nodes. This is due to the restrictions on food availability and also waste disposal at the nodes. Using a smaller number of bacteria limits the maximum range of the molecular concentration output by the transmitter. Therefore, in this paper, we resort to use relaying to mitigate this problem. We show that relaying can result in either increasing the maximum range of the concentration of molecules at the receiver or diversifying the output. This effectively results in higher capacity and communication reliability; as if we have increased the effective number of bacteria in the nodes.

Relaying and multi-user molecular communication have been previously discussed briefly in some other contexts. The design of repeaters in Calcium junction channels is discussed in~\cite{Nakano_Repeater}. Authors in~\cite{Atakan_Multi} considered the multi-user problem in molecular communication and compared it to its conventional counterparts. In contrast, in this paper, we analyze the relay problem in a practical molecular communication network that involves bacteria as fundamental agents in the nodes for signal transmission and reception. We consider different relaying cases and show how the information rate is improved by using the relay node.

The rest of the paper is as follows: in the Sec.~\ref{sec:review}, we briefly review the two-node molecular communication. Sec.~\ref{sec:simple_relay} introduces two scenarios for sense and forward relaying. The decode and forward relaying in M-ary signaling is discussed in~\ref{sec:Decode_Forward}. Finally, Sec.~\ref{sec:conclusion} concludes the paper.


\section{Two-node Molecular Communication}
\label{sec:review}

\begin{figure}
\centering
\vspace{-.2in}
\includegraphics[width= .4\textwidth]{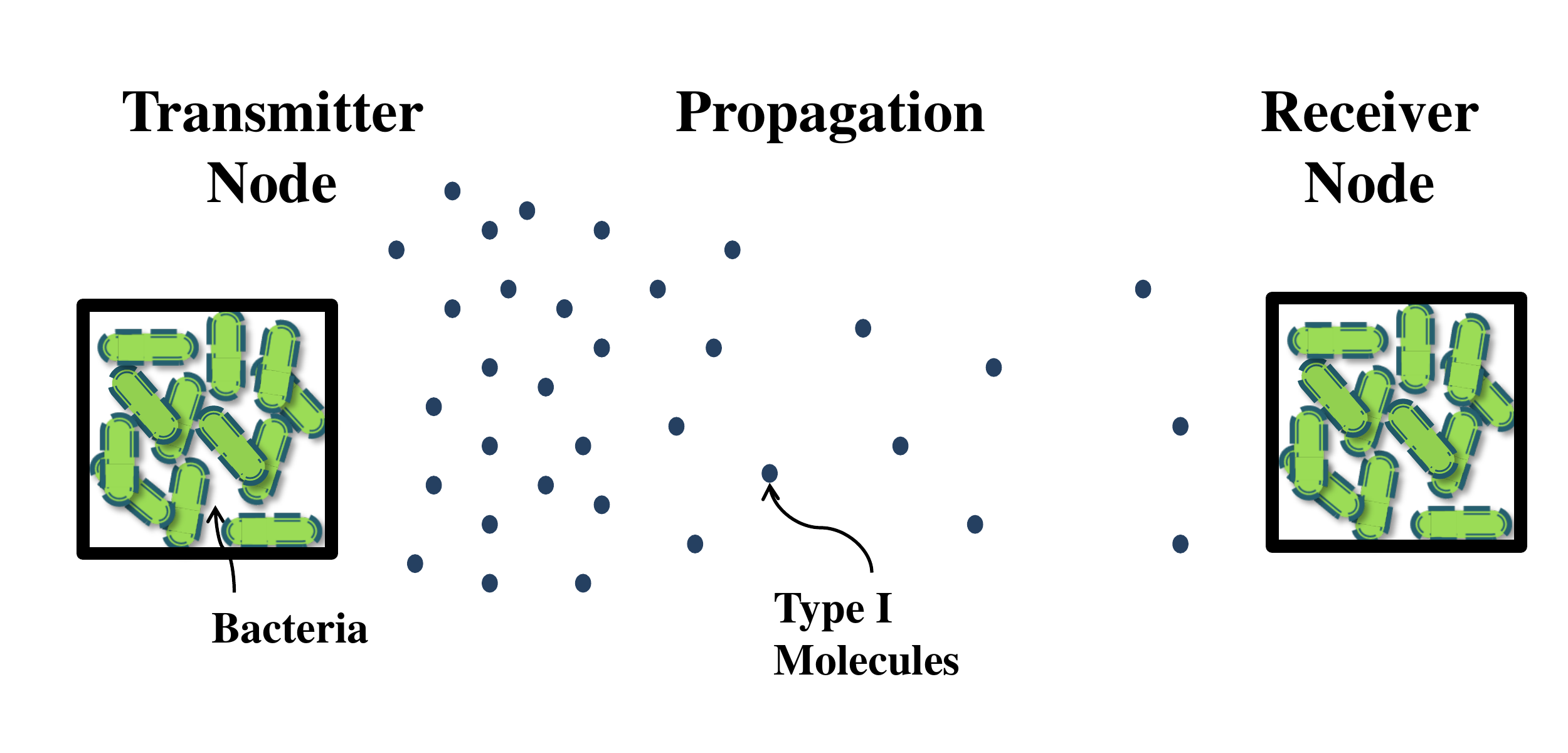}
\vspace{-.2in}
\caption{Molecular communication setup consisting of the transmitter, channel and the receiver}
\vspace{-.1in}
\label{fig:two_node}
\end{figure}

In this section, we briefly review the model we use for molecular communication between a single transmitter and receiver which will be used for the relaying scenario later on. Here, the information is encoded in the level of concentration of molecules that the transmitter can induce at the receiver node. As we can see in Fig.~\ref{fig:two_node}, upon stimulation, bacteria at the transmitter node produce signal molecules (namely type I) which are in turn diffused freely into the environment. The stimulation of the transmitter bacteria can be from the chamber or in reaction to other nodes in the network which is discussed in~\cite{INFOCOM2012_Arash}. Here, we assume that the transmitter node is able to produce continuous levels of concentration of molecules in the range $[0, A_{max}]$ at the receiver where the maximum level $A_{max}$ depends on the transmitter functionality.

Upon the reception of molecules at the receiver, as in Fig.~\ref{fig:two_node}, several binding events may occur between the received molecules and receptors in the bacteria~\cite{muller2008}. As such, a chain of chemical processes is triggered at each individual bacterium of the receiver node. The final output of each bacterium, once the chemical processes are over, is determined by the type of synthetic bacteria and is programmed into its plasmid. The output can be in the form of fluorescent (in particular GFP), the production of another type of molecules and so on. Here, the output of the receiver node is considered to be GFP resulting from all bacteria in the node. The level of the output GFP depends on the level of concentration of molecules the bacteria sense at the receiver. Ideally, by measuring this concentration, we can decode the transmitter information. As we have shown in~\cite{ITW2012_Arash}, the uncertainty in the communication is due to the discrepancy in the behavior of bacteria in a population. It was also shown that the dominant factor in the noise (i.e., uncertainty), comes from the probabilistic nature of the molecular signal sensing by the bacteria at the receiver. We denote by $A_0$ the concentration of the received molecules and by $p_0$ the ideal probability of the activation of individual receptors in bacteria due to molecular binding. Using~\cite{muller2008}, we have
\begin{equation}
\label{eq:steady_state}
p_0=\frac{A_0\gamma}{A_0 \gamma+\kappa},
\end{equation}
where $\gamma$ and $\kappa$ are some parameters due to synthetic bacteria. They can be viewed as random variables who vary around their averages by zero-mean Gaussian noises $\epsilon_{\gamma}$ and $\epsilon_{\kappa}$ with variances  $\sigma_{\gamma}^2$ and $\sigma_{\kappa}^2$, respectively. The noisy probability of the activation is obtained by~\cite{ITW2012_Arash} as
\begin{equation}
\label{noisy_probability}
p=p_0+p_0(1-p_0)(\frac{\epsilon_{\gamma}}{\gamma}+\frac{\epsilon_{\kappa}}{\kappa}).
\end{equation}
 The final output of the receiver proportionally depends on $Y$ the number of activated receptors of all the bacteria in the population. We have~\cite{ITW2012_Arash}:
\begin{equation}
\left\{
\begin{array}{rcl}
\mathbf{E}[Y]&=& nNp_0\\
\mathbf{Var}(Y) &= &nN^2 p_0^2(1-p_0)^2 \sigma_0^2.
\end{array}
\right.
\label{eq:sink_node}
\end{equation}
where $n$ is the number of bacteria in each node, $N$ is the number of receptors in each individual bacterium and $\sigma_0^2= \frac{\sigma_{\gamma}^2}{\gamma^2}+ \frac{\sigma_{\kappa}^2}{\kappa^2}$. It is evident from~(\ref{eq:steady_state}) that $p_0$ can be viewed as the signal, since by obtaining/decoding $p_0$, we can infer the transmitted concentration (information). As we note from~(\ref{eq:sink_node}), the variance of the output depends on the signal (i.e., $p_0$) and takes its maximum at $p_0=\frac{1}{2}$. 
Finally, we note that if the rate of production of molecules at the transmitter output is $\alpha$, due to the channel diffusion, the steady-state concentration at distance $r$ from the transmitter is given by~\cite{random_walk}
\begin{equation}
\label{eq:concentration_received}
A_r=\frac{\alpha }{4\pi D r}.
\end{equation}
where $D$ is the channel diffusion coefficient.

%

We analyzed the information rate and probability of error for such a setup in~\cite{ITW2012_Arash} and showed how the capacity increases with increasing $A_{max}$ or $n$. In the next section, we study how a simple relay can increase the information rate (and reduce the probability of error) by influencing $A_{max}$ or the effective $n$.


\section{Sense and Forward Relaying}
\label{sec:simple_relay}

\begin{figure}
\centering
\vspace{-.12in}
\includegraphics[width= .35\textwidth]{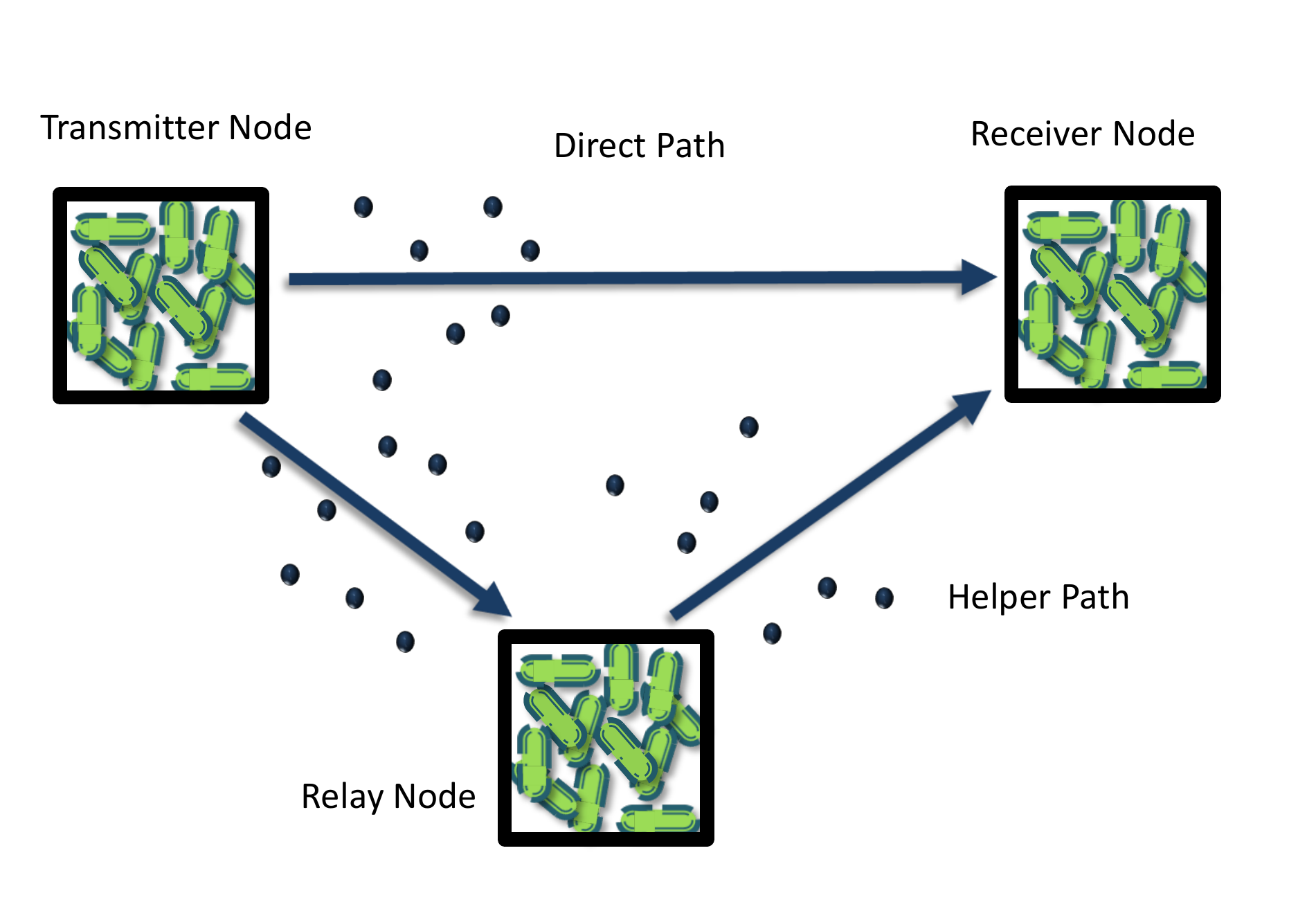}
\vspace{-.2in}
\caption{Communication via molecular relaying}
\vspace{-.15in}
\label{fig:relay}
\end{figure}

The schematic for a relay setup is shown in Fig.~\ref{fig:relay}. In this setup, the relay node forms another path to the receiver node to help the receiver in decoding the information. We assume nodes are able to produce any continuous concentration of molecules in their maximum range. Moreover, no decoding occurs at the relay node and the same concentration (or a constant multiple of it) is relayed to the receiver. In the setup shown in Fig.~\ref{fig:relay}, the direct distances from the transmitter to the receiver, from the transmitter to the relay and from the relay to the receiver are assumed $r_1$, $r_2$ and $r_3$, respectively. 

Since the movement of molecules in the diffusion channel and the process of the reception and production of molecules result in huge amounts of delay, naturally the molecules produced by the relay node reach the receiver after the reception of the molecules from the transmitter node. In other words, the direct-path output is always obtained prior to the relayed molecules reaching the receiver. In this paper, we assume the receiver is not able to store the signal form the transmitter in order to use it later with the signal from the relay information for better decoding of the information.

Assume that the transmitter produces the concentration $A_0$ at the receiver node.  Hence, using~(\ref{eq:sink_node}), the output $Y_1$ at the receiver due to the direct transmission would be 
\begin{equation}
Y_1=nNp_0+\epsilon_1
\end{equation}
where $\epsilon_1$ is a zero-mean Gaussian noise with signal-dependent variance given in~(\ref{eq:sink_node}).
Note that the diffusion channel has a broadcast nature. Therefore, the output molecular concentration of the transmitter node travels to both the receiver and relay nodes. Since the steady-state concentration of molecules depends inversely on the distance as shown in~(\ref{eq:concentration_received}), the average concentration of molecules received at the relay would be $A_2=\frac{A_0r_1}{r_2}$. In response to the molecular concentration $A_2$, similar to the receiver node in Sec.~\ref{sec:review}, the relay would produce $Y_2$ as
\begin{equation}
Y_2=nNp_2+\epsilon_2
\end{equation} 
where $p_2=\frac{A_2\gamma}{A_2 \gamma+\kappa}$ and $\epsilon_2$ is a signal-dependent zero-mean Gaussian noise whose variance can be obtained from~(\ref{eq:sink_node}) by replacing $p_0$ with $p_2$. That is $
\mathbf{Var}(Y_2) = nN^2 p_2^2(1-p_2)^2 \sigma_0^2.$

Being stimulated by the type I molecules from the transmitter, the output of the relay node which depends on $p_2$, goes through a nonlinear transform (performed by the node) in order to induce the concentration $A_3=\beta A_2$ of molecules at the receiver. Here, $\beta$ is a constant amplification factor due to the chemical processes. We refer to this scenario as the sense (the concentration $A_2$) and forward relaying. Whether the molecules used for this relaying are the same as the transmitter node's molecules (i.e., type I) or another type of molecules (namely type II), the problem is divided to two cases. In the following, we study and compare both cases. Note that throughout the paper, it is assumed that the necessary relative distances are known to the corresponding nodes. These distances can be programmed into the nodes or estimated during a training phase.


\subsection {Single-type Molecule Relaying}

Here, the relay produces type I molecules. We assume that the nodes have the same maximum rate of type I molecule production and hence, $\beta=\frac{r_2}{r_3}$. Therefore, the total steady-state concentration of molecules $A_R$ at the receiver is equal to the aggregate of molecules received from both the transmitter node and the relay. That is:
\begin{equation}
\label{eq:concentration_receiver}
A_R=A_0+A_0 \frac{r_1}{r_3}+\epsilon_r
\end{equation}
where $\epsilon_r$ is a zero-mean noise due to the reception process at the relaying (i.e., a function of $\epsilon_2$ and distance $r_3$).

It was shown in~\cite{ITW2012_Arash} that due to the low-pass nature of the chemical processes inside the bacteria, i.e.,  the averaging operations that happen due to the collective reception of molecules by the bacteria, as well as low-pass nature of diffusion process, the last stage noise due to the the probabilistic reception of molecules by the receiver dominates the other noises (e.g., the transmitter noise) accumulated in the previous stages. Since no decoding is done by the relay, the noise resulted from the relay is dominated by the reception process at the receiver as in the case of the transmitter noise in~\cite{ITW2012_Arash}.

Thus the concentration of molecules $A_R$ sensed by the receiver produces the output (in form of GFP) $Y_R$ given by 
\begin{equation}
\label{eq:concentration_receiver}
Y_R=nN p_R+\epsilon_R
\end{equation}
where $p_R=\frac{A_0 (1+\frac{r_1}{r_3})\gamma}{A_0 (1+\frac{r_1}{r_3}) \gamma+\kappa}$ and $\epsilon_R$ is a signal dependent zero-mean Gaussian noise with variance $nN^2 p_R^2(1-p_R)^2 \sigma_0^2$. Note that $\epsilon_R$ is the noise only due to the molecular reception process at the receiver.

\begin{figure}
\vspace{-.4in}
\includegraphics[width= .4\textwidth]{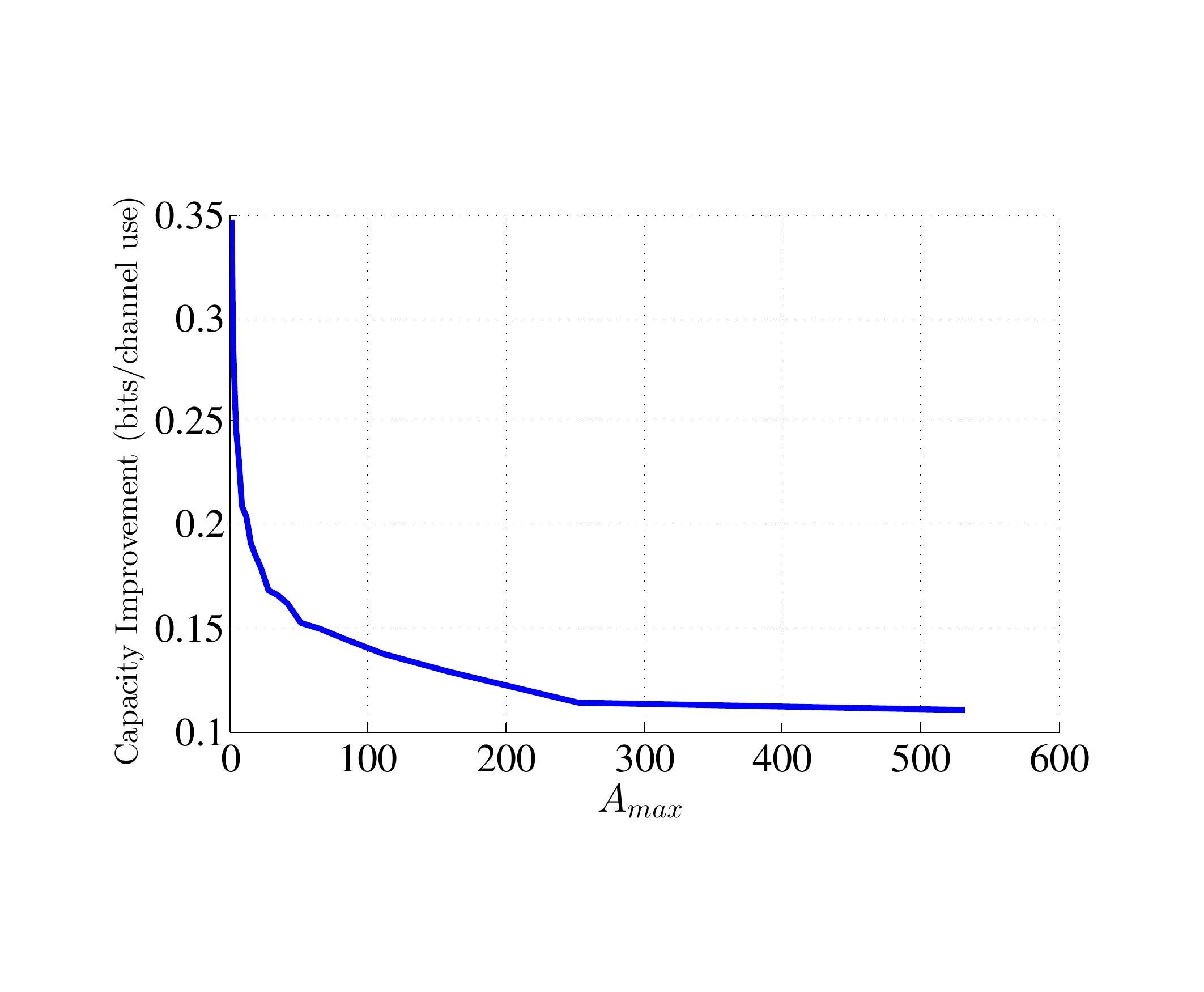}
\vspace{-.4in}
\caption{Capacity improvement in sense and forward relaying with a single molecule type}
\label{fig:relay_one_type}
\end{figure}

Using~(\ref{eq:concentration_receiver}), the capacity of the system involving relay can be computed as in~\cite{ITW2012_Arash} by maximizing $I(p_R;Y_R)=I(A_0;Y_R)=I(p_0;Y_R)$. 
 In~\cite{ITW2012_Arash}, we showed that one major factor in the capacity of molecular communication is the range of concentration of molecules $[0\; A_{max}]$ that the transmitter can induce at the receiver. As we see in~(\ref{eq:concentration_receiver}), the relaying in this case resulted in increasing $A_{max}$ to $A_{max}(1+\frac{r_1}{r_3})$. Note that $A_{max}$ is inversely proportional to $r_1$. Moreover, in order to be able to neglect the relay noise, $r_3$ should be large enough (compared to $r_2$). In Fig.~\ref{fig:relay_one_type}, we have shown the improvement in the capacity resulted by relaying for different values of $A_{max}$. Here, we assumed $n=25$, $N=10$ and $r_1=r_2=r_3$. Since the computations are prohibitive for large $n$ and $N$, we use small values for $N$ and $n$. As shown by the plot, the effect of relaying decreases for large $A_{max}$.

\subsection{Multi-type Molecule Relaying}

In this scenario, the relay uses another type of molecules, namely type II, in response to receiving type I molecules from the transmitter. Hence, there will be two types of molecules at the receiver which result in two different outputs, e.g., GFP and YFP, the Green and Yellow Fluorescent Proteins. Here, we neglect the interference between the two different types of molecules at the receiver and assume they act independently. The rationale is that the reception and production of molecules at relay incur huge amount of delay which translates into different reception times for type I and II molecules at the receiver.

We assume that the reception process is basically the same for the two types of molecules but each type has its own separate receptors in the bacteria. The output of each bacteria is directly related to the number of activated receptors. Since the output is different for each type, we consider different constants for the two types. Note that in the previous section, we did not explicitly consider this constant since there was only one type of molecules. Hence, the two outputs can be written as 

\begin{equation}
\left\{
\begin{array}{rcl}
Y1=\alpha_1 nNp_0+\epsilon_1\\
Y2=\alpha_2 nNp'_0 +\epsilon_2
\end{array}
\right.
\label{eq:two_output}
\end{equation}
where $p'_0=\frac{\beta A_0 \frac{r_1}{r_2}\gamma}{\beta A_0 \frac{r_1}{r_2}\gamma+\kappa}$. In addition, $\epsilon_1$ and $\epsilon_2$ are zero-mean noises with variances $n N^2 \alpha_1^2 p_0^2(1-p_0)^2 \sigma_0^2$ and  $n N^2 \alpha_2^2{ p'}_0^2 (1-p'_0)^2 \sigma_0^2$, respectively. Here, we have assumed that the number of receptors for the two types are equal.

In order to make the analysis tractable, we assume the amplification factor in production of type II molecules is $\beta=\frac{r_2}{r_1}$. Hence, we have $p'_0=p_0$. The goal is to maximize the mutual information between the input $p_0$ and the outputs $Y_1$ and $Y_2$:
\begin{equation}
C= \text{max}_{p_0} I(p_0;Y_1,Y_2)=I(p_0;\hat{Y_1},\hat{Y_2})
\label{eq:capacity_joint}
\end{equation} 
where $\hat{Y_1}=\frac{Y_1}{\alpha_1}$ and $\hat{Y_2}=\frac{Y_2}{\alpha_2}$.

 The equations in~(\ref{eq:two_output}) resembles the ones in case of Single-Input/Multiple Output (SIMO) configuration. With constant variance noises, the sum of two outputs $\hat{Y}_1$ and $\hat{Y}_2$, i.e., performing Maximum Ratio Combining, would have been the sufficient statistics for decoding $p_0$:
\begin{equation}
\label{eq:sum_outputs}
\hat{Y}=2nNp_0+\epsilon
\end{equation}
where $\epsilon$ is a zero-mean Gaussian noise with variance $2nN^2 p_0^2(1-p_0)^2 \sigma_0^2$. In our case, however, since the noise terms are only conditionally Gaussian, $Y$ is not the sufficient statistics for detecting $p_0$. As such, some information would be lost if we use only $\hat{Y}$. The output $Y_1$ and $Y_2$ are independent given $p_0$, i.e., $P(Y_1,Y_2|p_0)=P(Y_1|p_0)P(Y_2|p_0)$. Hence, we can use the numerical Blahut Arimoto algorithm for the joint outputs to maximize the joint mutual information in~(\ref{eq:capacity_joint}). 

\begin{figure}
\vspace{-.5in}
\includegraphics[width= .4\textwidth]{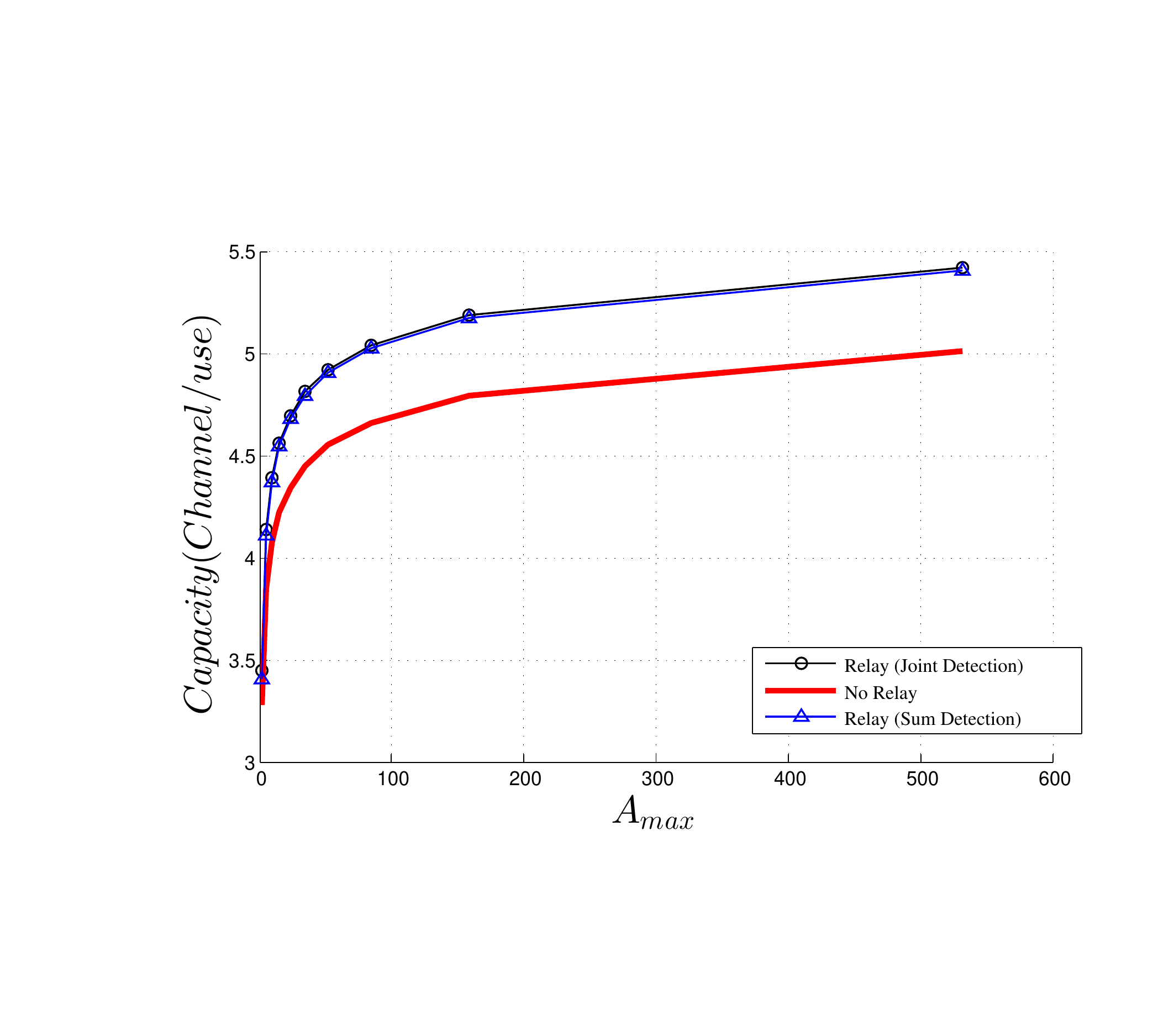}
\vspace{-.5in}
\caption{The capacity improvement by using two types of molecules in sense and forward relaying}
\label{fig:amplify_relay}
\end{figure}

In Fig.~\ref{fig:amplify_relay}, we have shown the numerical results for the maximum mutual information $I(p_0;\hat{Y}_1,\hat{Y}_2)$ as well as $I(p_0;\hat{Y})$. The case without the relay is shown for comparison as well. Equation~(\ref{eq:sum_outputs}) implies that the mutual information is increased by (2n) in relaying versus (n) without relaying. The results are shown for $\sigma_0^2=0.1$, $n=25$ and $N_1=N_2=10$. 

As we see in the plots, the difference between using the sum $\hat{Y}$ instead of both $\hat{Y}_1$ and $\hat{Y}_2$ to decode $p_0$ is almost negligible. In other words, the improvement of using two different outputs (e.g., GFP and YFP) instead of using only one type of output which is resulted from aggregation of type I and type II molecules output is small.  In~\cite{ITW2012_Arash}, we showed as to how the mutual information increases with increasing $n$. Finally, we note from Fig.~\ref{fig:amplify_relay} that the capacity improvement due to multi-type molecular relay is comparable to the single-type relaying in low concentration ranges but outperforms it for high concentration ranges.

\section{Decode and Forward Relaying in M-ary Signaling}
\label{sec:Decode_Forward}

In the previous section, we discussed the cases that the relay forwards a continuous concentration of molecules that it senses. Here, we consider a more practical scenario in which the transmitter uses M-ary signaling to communicate with the receiver as in~\cite{ITW2012_Arash}, however, with the help of relay. Here, we consider the case that the transmitter broadcasts the type I molecules for an M-ary symbol to both the receiver and the relay. Upon the reception and decoding the transmitted M-ary symbol, the relay node transmits the same symbol to the receiver with the same type I molecules. The aggregate of molecules arrived from the direct path and from the relay in Fig.~\ref{fig:relay}, are used to decode the symbols at the receiver.

We obtained the optimal distribution of the symbols for a direct communication from the transmitter to the receiver in~\cite{ITW2012_Arash}. We showed that due to the noise structure, larger weights should be given to the symbols assigned to either large or low concentrations. For symbol $i$, $(0\leq i \leq M-1 )$, let $A_i$ and $B_i$ be the concentrations of molecules induced by the transmitter at the receiver node and the relay node, respectively. Likewise, let $C_i$ be the concentration induced by the relay at the receiver node. As discussed in the previous section, by the broadcast nature of the diffusion channel, producing the symbol $A_i$ at the receiver corresponds to the production of the concentration $B_i=A_i \frac{r_1}{r_2}$ at the relay. In other words, each symbol $A_i$ designed optimally in the range $[0\;A_{max}]$ for the receiver node is mapped linearly to the symbol $B_i$ in $[0\; \frac{r_1}{r_2}A_{max}]$ at the relay node. Hence, the distribution of the $B_i$ symbol at the relay input would be optimal as well. 

\begin{figure}
\centering
\vspace{-.5in}
\includegraphics[width= .4\textwidth]{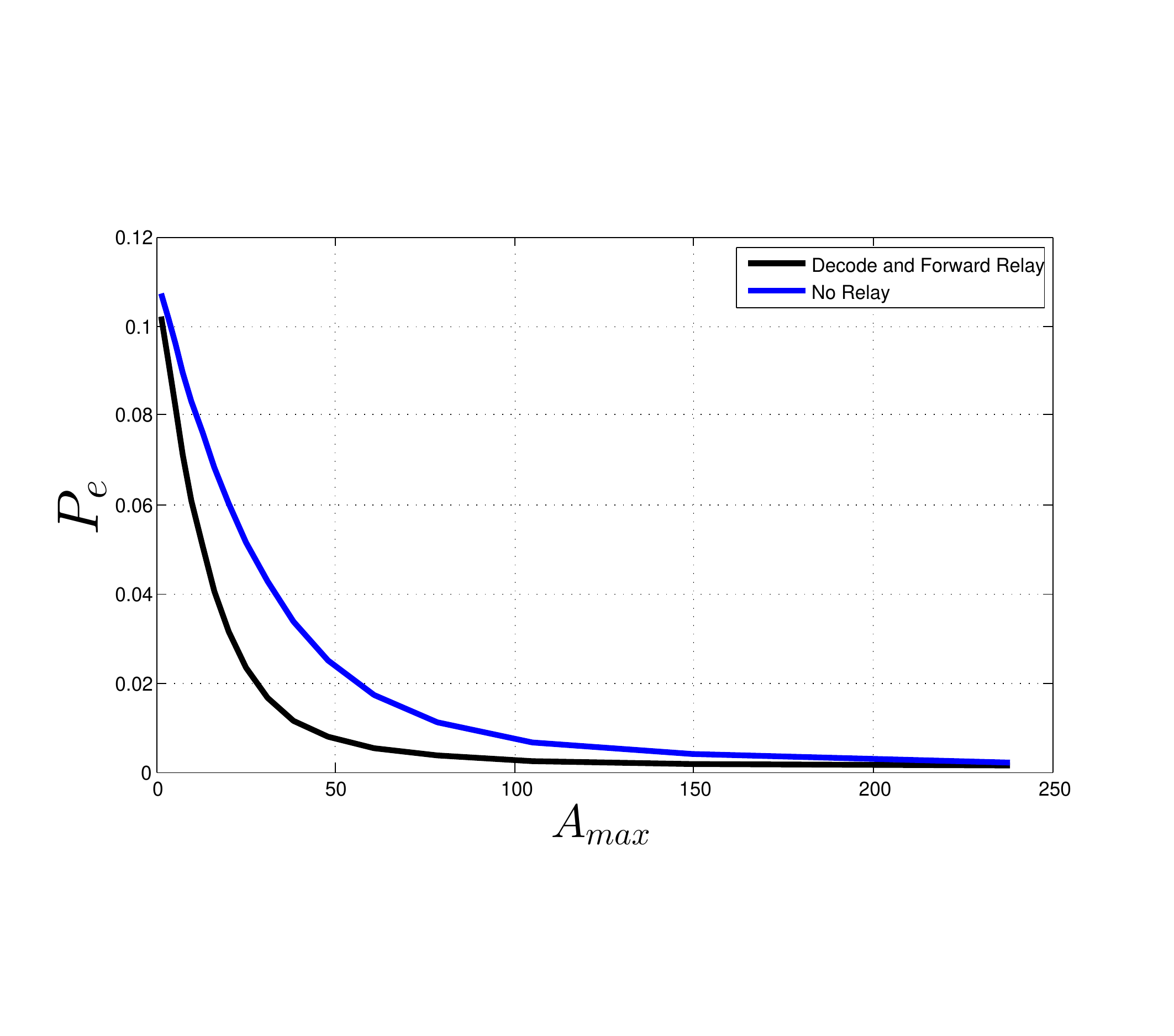}
\vspace{-.5in}
\caption{Probability of error with and without use of decode and forward relaying for M=8}
\vspace{-.21in}
\label{fig:mary_relay}
\end{figure}

Assume the symbol $B_j, (0\leq j \leq M-1 )$, is decoded at the relay. In response, the relay induces the symbol $C_j$ in the range of $[0\;\frac{r_1}{r_3}A_{max}]$ at the receiver. Then the aggregated concentration of molecules from both the direct and relay paths would be $A_i+C_j$ which is in the range $[0\;A_{max}(1+\frac{r_1}{r_3})]$. The output $Y$ at the receiver is given by
\begin{equation}
Y=\frac{(A_i+C_j)\gamma}{(A_i+C_j)\gamma+\kappa} +\epsilon,
\label{eq:receiver_relay_discrete}
\end{equation}   
where $\epsilon$ is a signal-dependent noise as in the previous section. We assume the receiver uses maximum a posteriori (MAP) detection to decode the symbol $A_i$. Hence, we have
\begin{align}
P(A_i|Y)&\propto P(Y|A_i)P(A_i)\nonumber\\
&=\sum_{j=0}^{M-1} {P(Y|A_i,C_j)P(C_j|A_i)P(A_i)},
\label{eq:receiver_max}
\end{align}
In~(\ref{eq:receiver_max}), $P(Y|A_i,C_j)=P(Y|A_i+C_j)=P(\epsilon|A_i+C_j)$ where $\epsilon$ is the conditionally Gaussian noise in~(\ref{eq:receiver_relay_discrete}).  Moreover,  since the mapping from $B_j$ to $C_j$ and from $A_i$ to $B_i$ are one-to-one, $P(C_j|A_i)=P(B_j|B_i)$ and we have
\begin{equation}
P(B_j|B_i)=P(j=arg\text{max}_k P(B_k|Y_2))
\label{eq:relay_max}
\end{equation}
 where $Y_2$ is the output at the relay with a conditionally Gaussian distribution $ \mathbf{N}(nNp_i,p_i^2(1-p_i)^2 \sigma_0^2)$ in which $p_i=\frac{B_i\gamma}{B_i\gamma+\kappa}$. By solving the optimization problems in~(\ref{eq:receiver_max}) and~(\ref{eq:relay_max}) simultaneously, the receiver can detect the input. In Fig.~\ref{fig:mary_relay}, we have shown the probability of error resulted from this scheme and compared it to the the case without relay for $M=8$. The results are given for $N=n=50$, $\sigma_0^2=0.1$ and $r_1=r_2=r_3$. As we see in Fig.~\ref{fig:mary_relay}, relaying reduces the probability for error but this improvement vanishes for high ranges of concentration.

\vspace{-.1in}
\section{Conclusion}
\label{sec:conclusion}

In this paper, we studied molecular communication relaying. We considered both the case that the relay node simply senses and forwards the received continuous concentration of molecules and also the case that relay decodes the received M-ary symbol and forwards it to the receiver. In the first case, we showed that if the relay uses the same type of molecules as the transmitter, relaying effectively results in expanding the range of the concentration of molecules at the receiver, and hence, increases the capacity. This effect is more significant for the low range of concentrations. On the other hand, by using a different type of molecules, relaying nearly corresponds to increasing the effective number of bacteria in the nodes which increases the capacity for all ranges of low and high concentrations. For decode and forward relaying with M-ary scheme, we assumed all the nodes use the same type of molecules. We showed that by using MAP decision making, the probability of error decreases. Moreover, this improvement decreases for higher ranges of concentration of molecules at the receiver.

\vspace{-.15in}
\bibliographystyle{IEEEtran}
\bibliography{ISIT2013}
\end{document}